\documentstyle[aps,psfig]{revtex}
\begin{document}
\title{Chiral symmetry restoration and the string picture of hadrons.}
\author{ L. Ya. Glozman}
\address{  Institute for Theoretical
Physics, University of Graz, Universit\"atsplatz 5, A-8010
Graz, Austria\footnote{e-mail: leonid.glozman@uni-graz.at}}
\maketitle

\begin{abstract} 
QCD string picture of highly excited hadrons very
naturally explains parity doubling once the chiral
symmetry is restored high in the spectrum. In particular,
the spin-orbit and tensor interactions of
quarks at the ends of the string, related to dynamics of 
the string, vanish. High in the spectrum there appears
higher degree of degeneracy, namely parity doublets
with different angular momentum cluster around energy
of the string in the given quantum state.
\end{abstract}

\bigskip
\bigskip

It has recently been shown that  the validity of the
operator product expansion in QCD at large space-like
momenta and the validity of the dispersion relation for
the two-point correlator (i.e. validity of the K\"allen -
Lehmann representation) implies a smooth chiral symmetry
restoration in  the high-energy part of hadron spectra composed of
light quarks \cite{CG1,CG2}. Indeed, at large space-like
momenta $Q^2$ the difference between the two correlators
$\Pi_{J_1}(Q^2)$ and $\Pi_{J_2}(Q^2)$ obtained with the
local currents $J_1$ and $J_2$ which are connected by the
chiral transformation $J_1 = UJ_2U^\dagger$,
$U \in SU(2)_L \times SU(2)_R$
 is due only to
the small contributions of the chiral condensates that are
suppressed by inverse powers of momenta, 

\begin{equation}
\Pi_{J_1}(Q^2) - \Pi_{J_2}(Q^2) \sim \frac{1}{Q^n}, ~ n > 0.
\end{equation}

\noindent
Since the large $Q^2$ correlator is completely dominated by the
large $s$ spectral function, one  expects that at large
$s$ the theoretical spectral functions $\rho_{J_1}(s)$ and 
$\rho_{J_2}(s)$ must become essentially the same. This manifests
a smooth chiral symmetry restoration from the low $s$ region,
where both spectral functions are very different because of the
chiral symmetry breaking in the vacuum,  to the high $s$ region,
where the chiral symmetry breaking in the vacuum becomes irrelevant.\\

Microscopically this is because
the typical momenta of valence quarks should increase
higher in the spectrum and once it is high enough the
valence quarks decouple from the chiral condensates of
the QCD vacuum and the dynamical (quasiparticle or constituent)
mass of quarks drops off and the chiral symmetry gets restored
\cite{G}. This phenomenon does not mean that the spontaneous
breaking of chiral symmetry in the QCD vacuum  disappears,
but rather that the chiral asymmetry of the vacuum becomes
irrelevant  sufficiently high in the spectrum. While
the chiral symmetry breaking condensates in the vacuum are
crucially important for the physics of the low-lying hadrons,
the physics of the highly-excited states is such as if there
were no chiral symmetry breaking in the vacuum.\\

This is in a very good analogy with the similar phenomenon
in condensed matter physics. Namely, even if the metal is
in the superconducting phase (i.e. there is a condensation of
the Cooper pairs - which is analogous to the condensation of right-left
quark pairs in the QCD vacuum - and the low-lying excitations of the system
are the excitations of quasiparticles), the response of the superconductor
to the high-energy (frequency) probe  $\hbar \omega >> \Delta$
is the same as of the normal metal. This is because the phase
coherent effects of the superconductor become irrelevant in this case
and the external probe sees a bare particle rather than a quasiparticle.
Similar, in the QCD case one has to probe the QCD vacuum by the
high energy (frequency) external probe (current) in order to create a hadron
of a large mass and hence the masses (physics) of the 
highly excited hadrons should
be insensitive to the condensation of chiral pairs in the vacuum.\\

If high in the spectrum (i.e. where the chiral symmetry is
approximately restored) the spectrum is still quasidiscrete, then
the phenomenological manifestation of the chiral symmetry restoration
would be that the highly excited hadrons should fall into the representations
of the $SU(2)_L \times SU(2)_R$ group, which are compatible with the
definite parity of the states - the parity-chiral multiplets \cite{CG1,CG2}.
In the case of baryons in the $N$ and $\Delta$ spectra these
multiplets are either the parity doublets ($(1/2,0) \oplus (0,1/2)$
for $N^*$ and  $(3/2,0) \oplus (0,3/2)$ for $\Delta^*$) that are
not related to each other, or the multiplets $(1/2,1) \oplus (1,1/2)$
that combine one parity doublet in the nucleon spectrum with the
parity doublet in the delta spectrum with the same spin. And indeed,
the phenomenological highly-excited $N$ and $\Delta$ spectra show
systematic parity doubling \cite{CG1,CG2}, 
though further experimental studies
are required in order to make a definitive statement. It remains
to be clarified which  parity-chiral multiplets are
observed - at the moment one cannot disentangle the different possibilities.
In Fig. 1 we show the  experimental $N$ spectrum \cite{PDG}.
Starting from $M \simeq 1.7$ GeV it is clear that the states
systematically appear as approximate parity doublets, except for
the two high-lying high-spin states, where the chiral partner
have yet to be experimentally found. The onset of the approximate
chiral symmetry restoration as low as in the region of 1.7 GeV
is very consistent with the recent finding of JLAB that the
inclusive electroproduction of resonances in the mass region
$3.1 \leq M^2 \leq 3.9$ GeV$^2$ is perfectly dual to 
deep inelastic data \cite{JLAB}.\\
\begin{figure}
\begin{center}
\centerline{\psfig{file=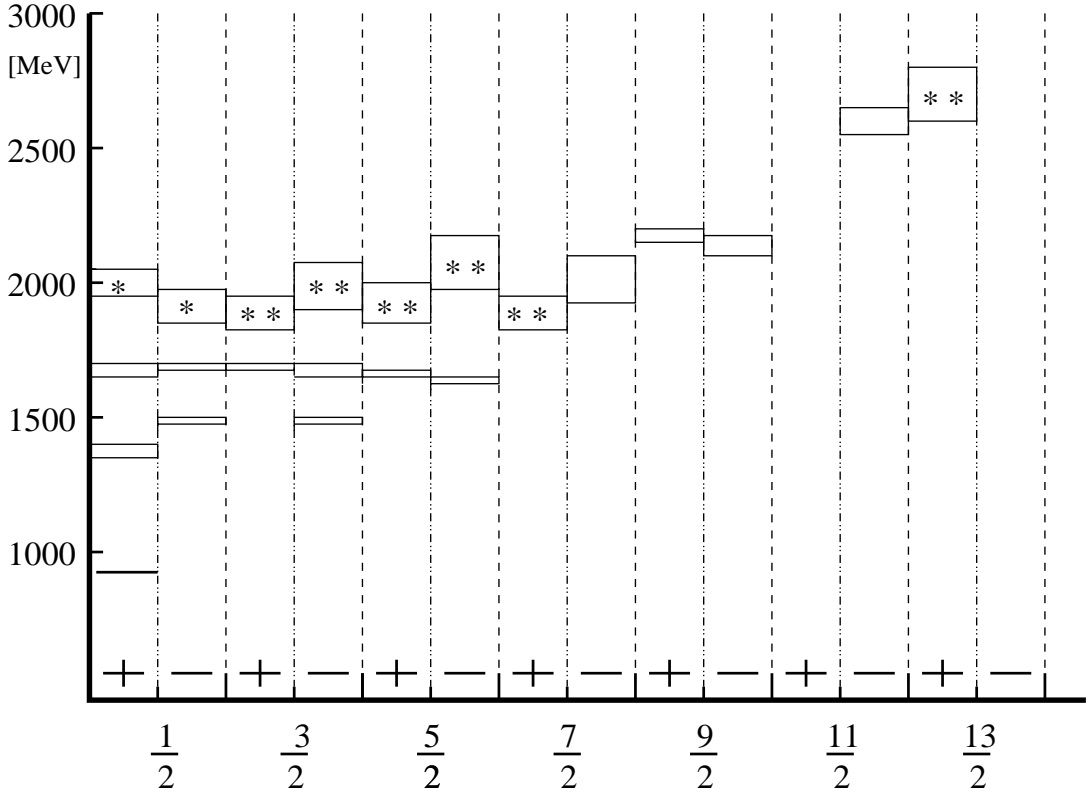,width=0.8\textwidth,angle=-90,clip=}}
\end{center}
\caption{Excitation spectrum of the nucleon. The real part of the pole
position is shown. Boxes represent experimental uncertainties.
Those resonances which are not yet reliably established are
marked by two or one stars according to the PDG classification.
The one-star resonances with $J=1/2$ around 2 GeV are given
according to the recent Bonn (SAPHIR) results [8].}
\end{figure}

Systematic data on highly-excited mesons are still missing
in the PDG tables. However, the data obtained from the ongoing partial wave
analysis of mesonic resonances in the proton-antiproton
annihilation at LEAR  \cite{BUGG} show an almost
 systematic appearance of  nearly perfect parity doublets
in the meson spectrum in the mass region of 1.8 GeV and above . 
A detailed analysis of the chiral multiplets with spin 0 is given
in \cite{AXIAL}. A similar situation is also observed for the highly 
excited mesons with higher spins.
 Yet the unique assignment of
these higher spin states to the specific multiplets is difficult
because sometimes the same state  fits perfectly into different
multiplets. For example, on the one hand
the $\rho(1,1^{--})$ states at $1970  \pm 30$
MeV and at $2265 \pm 40$ MeV can be paired with the
$a_1(1,1^{++})$ states at $1930^{+30}_{-70}$ MeV and 
$2270^{+55}_{-40}$, respectively ($(0,1) \oplus (1,0)$ representation). 
But one the
other hand the same $\rho$-s can  also be paired with $h_1(0,1^{+-})$
states at $1965  \pm 45$ and $2215  \pm 40$, respectively 
(i.e. the $(1/2,1/2)$ representation). Clearly a 
very systematic phase shift
analysis should be done for both baryon and meson spectra and
probably other properties such as decay modes should be invoked
in order to unambiguously identify possible chiral partners.\\

Obviously, one also needs a physical picture for highly excited
hadrons that is compatible with the chiral symmetry restoration
and which would naturally explain the parity doubling. It is a purpose
of this note to show that the QCD string picture of highly excited
hadrons   does this job.\\

It is well known that the string picture very naturally
incorporates the approximately linear behaviour of Regge
trajectories for excited states. Low in the spectrum,
where the spontaneous breaking of chiral symmetry in the vacuum
is crucially
important and which prevents the parity doubling, the linear
behaviour of Regge trajectories is heavily broken \cite{GREGGE}.
In this part of the spectrum the physics which is associated with
the quasiparticle picture (i.e. the constituent quark model) 
is more relevant and the effective
interactions of quasiparticles can be approximated to some
extent by potentials of different physical origin (e.g.
by effective confining potential, flavor-spin potential which
is associated with the Goldstone-boson-exchange between
constituent quarks in the low-lying baryons, etc.).\\

Consider, as an example,
the $\rho$ and $a_1$ mesons
 which become the chiral partners when
the chiral symmetry is restored. The spectra of both types of
mesons are extracted from the two-point functions

\begin{equation}
\Pi_{\mu \nu} =\imath \int d^4x ~e^{\imath q x}
\langle 0 | T \{ j_\mu (x) j^\dagger_\nu (0) \} |0\rangle,
\label{corr}
\end{equation}

\noindent
where the currents (interpolating fields) for the isovector-vector
and isovector-axialvector mesons are given as

\begin{equation}
 j^V_\mu (x) = \bar q(x) \gamma_\mu \frac{\vec \tau}{2} q(x) ,
\label{V}
\end{equation}

\begin{equation}
 j^{AV}_\mu (x) = \bar q(x) \gamma_\mu  \gamma_5 \frac{\vec \tau}{2} q(x).
\label{AV}
\end{equation}

The chiral $SU(2)_L \times SU(2)_R$ transformations consist of the isospin
 as well as the axial transformations. The isospin 
transformations leave the currents above invariant while the axial
transformations mix them

\begin{equation}
 j^V_\mu (x) \leftrightarrow j^{AV}_\mu (x).
\label{VAV}
\end{equation}

\noindent
Group theoretically both currents belong to the $(0,1) \oplus (1,0)$
representation of the parity-chiral group \cite{CG2}. \footnote
{$(I_L,I_R)$ denotes the irreducible representation of the chiral
group with $I_L$ and $I_R$ being the isospins of the left and
rigth quarks, respectively. Generally the irreducible representation of
the chiral group is not invariant under parity, which transforms
the left quarks into the right ones and vice versa. However, the
direct sum of two irreducible representations  $(I_L,I_R) \oplus (I_R,I_L)$
is invariant under parity and is an irreducible representation
of the parity-chiral group.}\\

In the chiral limit both the vector and the axial-vector currents are
conserved and the correlators for these currents can be written as

\begin{equation}
\Pi_{\mu \nu} = (q_\mu q_\nu  - q^2 g_{\mu \nu}) \Pi(q^2).
\label{corr2}
\end{equation}

\noindent
The imaginary part of $ \Pi(q^2)$ in the time-like domain
$s=q^2 > 0$ is proportional to the spectral function
$\rho(s)$. At large $s$ the spectral functions obtained with
the currents (\ref{V}) and (\ref{AV}) must coincide, 

\begin{equation}
\rho^V(s) = \rho^{AV}(s).
\label{rest}
\end{equation}

\noindent
This manifests the chiral symmetry restoration.
The theoretical spectral densities obtained in such a way can
probably be compared with the experimental ones only upon
some averaging \cite{WEIN}. For the present context the important
thing is that if at large $s$ the spectrum
is still quasidiscrete (i.e. it consists of separate
resonances), then the equality of the vector and axial vector
theoretical spectral densities is consistent with the approximate
degeneracy of the highly excited vector and axial vector mesons.
These mesons fill out in pairs the $(0,1)\oplus (1,0)$ representations
of the parity-chiral group.\\

Microscopically this means the following. {\it
At large} $s$ the valence quarks that
are injected by the current into the vacuum  do {\it not} become dressed by 
the quark condensates of the
vacuum and remain left or right on the whole way of their propagation
to the point where they are annihilated by the current. It is important
to construct a model picture for the highly excited hadrons
that is consistent with the given property of chiral symmetry restoration
and that would explain chiral multiplets high in the hadron spectra.\\

To carry out this task  we will assume for the highly excited hadrons
that  the QCD string-like picture of confinement
underlies the physics. We  make the following simplifying
assumptions: (i) the field in the string is of pure color-electric
origin;
(ii) the chiral symmetry is completely restored and hence
the hadrons are  rotating strings with {\it massless} quarks at
the ends with definite  {\it chirality=helicity} \footnote
{This is a short notation for $chirality = + helicity$ for quarks
and $chirality = - helicity$ for antiquarks.}
and these
valence quarks  are
combined into  parity-chiral multiplets.\\

While for mesons the geometry of the string is rather
obvious, it is not so for baryons.
In the present context it is not so important whether the
highly excited baryons are to be viewed as  symmetric
string configurations of the Mercedes-Benz type with the
string junction in the center or are of a deformed type
with the quark-diquark clustering. 
For us it is only important that the strings are pure
electric and that the valence quarks at the ends have definite 
chirality=helicity and belong to one of the parity-chiral multiplets.\\

\begin{figure}
\centerline{\psfig{file=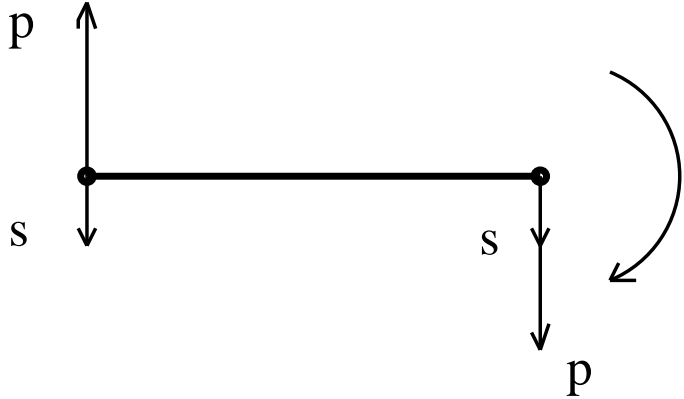,width=0.5\textwidth,angle=-90}}
\caption{ Rotating string with the right and the left quarks at the ends.}
\end{figure}

The motion of the quarks at the ends of the string
is constrained by the stationary motion of the string. This
motion is a rotation. The key point is that  
massless  $s=1/2$ 
quarks  in the chirally restored
regime possess a definite chirality=helicity (see Fig. 2). For
the description of the rotation of massless quarks with definite
chirality the angular momentum representation of the Weyl equations
must be used, which for the stationary motion 

\begin{equation}
\Phi_{R,L}(j,j_z,t) = \phi_{R,L}(j,j_z) e^{-\imath E t}
\label{ST}
\end{equation}

\noindent
are

\begin{equation}
\vec \sigma \cdot \vec \nabla \phi_{R}(j,j_z) = 
\imath E_R \phi_{R}(j,j_z),
\label{WR}
\end{equation}

\begin{equation}
-\vec \sigma \cdot \vec \nabla \phi_{L}(j,j_z) = 
\imath E_L \phi_{L}(j,j_z).
\label{WL}
\end{equation}

\noindent
Under parity transformation the Weyl equation for the
right spinor is transformed into Weyl equation for the
left spinor and vice versa. Since we require that the
valence quarks belong to the same parity-chiral multiplet,
then under parity operation the valence quarks must satisfy

\begin{equation}
 \phi_{L}(j,j_z) \leftrightarrow  \phi_{R}(j,j_z).
\label{PAR}
\end{equation}

\noindent
One then obtains that the rotating string does not distinguish
between the right and left valence quarks since they must have
the same energy

\begin{equation}
 E_{L} = E_{R}.
\label{EN}
\end{equation}

Since the nonperturbative field in the string is pure electric,
the magnetic interaction of quarks of the type 
$ \vec \sigma(i) \cdot \vec \sigma (j)$
is absent.
Though the string is pure electric, the quantum fluctuations of
perturbative origin are possible once the quarks are close 
to each other.\footnote{Depending on whether the nonperturbative
resummation of gluonic exchanges or the instanton-induced interactions
provide the chiral symmetry breaking in QCD, the corresponding forces
automatically  split the low-lying mesons
where chiral symmetry breaking effects are central to the physics.}
However, for massless quarks with definite chirality=helicity 
the interaction induced by perturbative forces, does not distinguish between 
the members of the parity-chiral multiplet.\\

The electric field in the string is "flavor-blind" and hence does not
distinguish between the light quarks of different flavor once the
chiral limit is taken.\\

Since chiral symmetry is assumed to be
restored, the valence quarks are decoupled from the Goldstone
bosons\footnote{The coupling of Goldstone bosons to quasiparticles
(constituent quarks) \cite{MG} is regulated by the 
Goldberger-Treiman relation taken at the quark level,
$g=\frac{M g_A}{f_\pi}$, where $M$ is the quasiparticle (dynamical)
mass
of quarks induced by the coupling of quarks to the quark condensates.
Chiral symmetry restoration high in the spectrum means (as it was
explained above)
that the valence quarks decouple from the quark condensates
which in turn implies that their quasiparticle mass $M$ approaches to the
current mass value (which is zero in the chiral limit). Hence the
 Goldstone bosons decouple from such valence quarks. This is
in contrast to the low-lying hadrons where the chiral symmetry
breaking effects are very important and where the coupling
of valence quarks to the chiral condensates is crucial for 
physics.}
 and hence the Goldstone boson mediated forces (e.g. the flavor-spin
 forces, that are crucially important in the
low-lying baryons \cite{GR}) between quarks are also absent.\\

One arrives at the following situation: (i) the hadrons with the
different chiral configurations of the quarks at the
ends of the string which belong to the same parity-chiral
multiplet
and that belong to the same intrinsic quantum state of the string
must be degenerate; (ii) the total parity of the hadron is determined
by the product of parity of the string in the given quantum state
and the parity
of the specific parity-chiral configuration of the quarks at the
ends of the string. There is no analogy to this situation in 
nonrelativistic physics where  parity is only determined by the 
orbital motion of particles.
Thus one sees that for every intrinsic quantum state of the string
there necessarily appears parity doubling of the states
with the same total angular momentum.\\

The spin-orbit operator $\vec \sigma \cdot \vec L$ does not
commute with the helicity operator $\vec \sigma \cdot \vec \nabla$.
Hence the spin-orbit interaction of quarks with the fixed
chirality=helicity is absent. In particular, this is also true
for the spin-orbit force due to the Thomas precession

\begin{equation}
U_T = -\vec \sigma \cdot \vec \omega_T \sim \vec \sigma
 \cdot [\vec v, \vec a]
\sim \vec v \cdot [\vec v, \vec a] =0,
\label{T}
\end{equation}

\noindent
where $U_T$ is the energy of the interaction and
$\vec \omega_T$, $\vec v$ and $\vec a$ are the angular frequency
of Thomas precession, velocity of the quark and its acceleration, respectively.\\

In addition, for the rotating string

\begin{equation}
\vec \sigma (i) \cdot \vec R(i)=0,
\label{T1}
\end{equation}

\begin{equation}
 \vec \sigma (i) \cdot \vec R(j)=0,
\label{T2}
\end{equation}

\noindent
where the indices $i,j$ label different quarks and $ \vec R$  
is the radius-vector of the
given quark in the center-of-mass frame.
 The relations above
immediately imply that the possible tensor interactions
of quarks related to the string dynamics should
be absent, once the chiral symmetry is restored.\\

As an example, consider a relativistic {\it potential} description of 
$\rho$ and
$a_1$ mesons within the constituent quark model \cite{GI}.
With this description the parity of the state is unambiguously
prescribed by the relative orbital angular momentum of the
quark and antiquark and the $\rho$ and
$a_1$ mesons are the $^3S_1$ and $^3P_1$ states, respectively.
 Clearly, such
a picture cannot explain the systematical degeneracy of
$\rho$ and $a_1$ mesons high in the spectrum because  stronger
centrifugial repulsion and weaker spin-spin force in $a_1$
mesons systematically shifts them  with respect to $\rho$
mesons. As a result, while the fitting of the parameters of
the model provides  success in the description of the {\it low-lying}
$\rho$ mesons, the model completely fails in the description
of $a_1$ mesons. Up to 2.4 GeV it predicts only two $a_1$ states (at
1.24 GeV and 1.82 GeV) while the experimental data indicates the states
at 1.26 GeV, 1.64 GeV, 1.93 GeV and 2.27 GeV \cite{BUGG}.\\

The failure of the potential description is inherently related to
its inability to incorporate chiral symmetry restoration. In contrast,
in the string picture once the chiral symmetry is restored
both $\rho$ and $a_1$ mesons belong in pairs to the {\it same} quantum
state of the string with the {\it same} angular momentum of the string.
The opposite parity of these states is provided by the different
right-left configurations of valence quarks.
While the vector mesons contain the following valence quark configuration 

\begin{equation}
\frac{1}{\sqrt 2} (\bar R \frac{\vec \tau}{2} \gamma^\mu R + 
\bar L \frac{\vec \tau}{2} \gamma^\mu L),
\end{equation}

\noindent
the axial vector mesons are constructed as

\begin{equation}
\frac{1}{\sqrt 2} (\bar R \frac{\vec \tau}{2}  \gamma^\mu R - 
\bar L \frac{\vec \tau}{2} \gamma^\mu L).
\end{equation}

\noindent
Both hadrons actually are the different parity states of the same
basic particle. This is like the $\rho^+,\rho^0, \rho^-$ which represent
 different charge (isospin) states of the same particle. While
 isospin symmetry (which is not broken in the vacuum) is a good
symmetry both for  low-lying and high-lying hadrons, 
chiral symmetry (which contains isospin as a subgroup) is
a good symmetry only high in the spectrum, in fact. So high in the spectrum
it would even be  proper to use the same "name" for chiral partners.
\\

As was shown above, in the string picture once the chiral 
symmetry is restored
 the spin-orbit
force is absent. Indeed, the
phenomenological meson and baryon spectra  in the $u,d$ sector
do not show large spin-orbit forces. In contrast, in the potential
picture the states are affected by the large spin-orbit force which
comes both from the one gluon exchange interaction  and the Thomas 
precession of quarks in the scalar confining potential. While for
some specific family of mesons the fine tuning of parameters can
provide an approximate cancellation of these  strong
spin-orbit interactions, this is not the case for the other family
of mesons. For example, though the  $^3S_1$ states in $\rho$-s are
not affected by the spin-orbit force, the spin-orbit force is strong
in $a_1$ mesons within this description.\\

 Once the quantum fluctuations of perturbative origin
are neglected, the energy of the hadron
is determined exclusively by the energy of the string;  the
mutial orientation of the spins of the quarks  become
irrelevant.
 In this case there appears an
even higher degree of degeneracy than the simple parity doubling.
Namely those hadrons with the same intrinsic energy and angular
momentum of the string but  with different total spin of all quarks
(that are allowed by Pauli principle) must be degenerate. 
The higher we are in the spectrum - the smaller is the amplitude for
the quarks to be close to each other and hence the perturbative
contributions become suppressed.
In other
words, one should expect an appearance of 
sets of degenerate parity-doublets with different total angular
momentum (i.e.  clusters of parity-doublets) in hadron spectra. For example,
if the intrinsic angular momentum of the string  in  an excited nucleon
is L=1, then
the cluster of parity doublets should consist of three doublets
with total angular momenta $J =1/2, 3/2, 5/2$ (because the total
spin of three quarks can range from 1/2 to 3/2). The first
cluster of parity doublets in the nucleon spectrum at $M \simeq 1.7$ GeV
neatly fits this picture. If one assumes that the next excitation
of the string is associated with  $L=2$, then there should be
cluster of parity doublets with $J=1/2,3/2,5/2,7/2$. And indeed
the band of baryons in the $N$ and $\Delta$ spectra in the mass
region $M  \simeq 2$ GeV does contain such parity doublets. The
empirical parity doublets in  meson spectrum with different total angular
momentum also perfectly group into two clusters around the
masses of $\simeq 2$ GeV and of $\simeq 2.2 - 2.3$ GeV.\\

Clearly, in reality the chiral symmetry restoration is only 
approximate and also the masses of light quarks are not zero,
 so the helicity is not equal to chirality 
and the quark spin can  flip with some small amplitude
during the rotation of the string. It can also be not parallel
to the quark momentum.
 Consequently there must appear 
weak spin-spin, spin-orbit and tensor interactions  of different 
origins between the valence quarks. These weak interactions should give 
rise to the small splittings in parity doublets as well as among
different doublets within the cluster. It is a challenging task
to evaluate such corrections.  

\bigskip

I am grateful to D.V. Bugg for providing me with data \cite{BUGG} on
highly excited meson spectra. This work was supported by the FWF
Project P14806-TPH.

\end{document}